\documentclass[a4paper,12pt]{article}

\usepackage[]{hyperref}
\usepackage{graphicx}
\usepackage{fancyheadings,color}
\pdfoutput=1
\usepackage{amssymb,amsmath,enumerate,mathrsfs,amsfonts}
\usepackage{graphicx,rotate,multicol}
\usepackage{multirow}
\usepackage{array}
\usepackage[numbers,sort&compress]{natbib}
\usepackage{blkarray}
\usepackage{multirow}
\usepackage{extarrows}
\usepackage{pdflscape}

\usepackage{dcolumn}% Align table columns on decimal point
\usepackage{bm}% bold math
\usepackage{dcolumn}% Align table columns on decimal point
\usepackage{textcomp}% Align table columns on decimal point
\usepackage{float}
\usepackage{subfig}
\usepackage{hypcap}
\usepackage{adjustbox}
\usepackage{makecell}
\usepackage{color}
\usepackage{pifont}
\usepackage{appendix}
\usepackage{url}
\usepackage{cancel}
\allowdisplaybreaks[1]
\usepackage{import}
\usepackage{subfiles}
\usepackage{filecontents}
\usepackage{slashed}
\usepackage{lineno}
\usepackage{placeins}
\usepackage{booktabs}

\let\Oldsection\section
\renewcommand{\section}{\FloatBarrier\Oldsection}

\let\Oldsubsection\subsection
\renewcommand{\subsection}{\FloatBarrier\Oldsubsection}

\let\Oldsubsubsection\subsubsection
\renewcommand{\subsubsection}{\FloatBarrier\Oldsubsubsection}
\hypersetup{
 unicode=false,          % non-Latin characters in Acrobat?s bookmarks
 pdftoolbar=true,        % show Acrobat?s toolbar?
 pdfmenubar=true,        % show Acrobat?s menu?
 pdffitwindow=true,     % window fit to page when opened
 pdfstartview={FitH},    % fits the width of the page to the window
 pdfsubject={GUT and Topological defects},   % subject of the document
 pdfnewwindow=true,      % links in new window
 pdfcreator={RevTeX},
 colorlinks=true,       % false: boxed links; true: colored links
 linkcolor=red,          % color of internal links
 citecolor=blue,        % color of links to bibliography
 filecolor=black,      % color of file links
 urlcolor=blue,           % color of external links
  }
  
\begin{document}

\title{
{ \begin{flushright}
\small \tt KW 19-009
\vspace*{0.5cm}
\end{flushright}}
Optimizing the Mellin-Barnes Approach to Numerical Multiloop Calculations 
\thanks{Presented by I. Dubovyk at the XLIII International Conference of Theoretical Physics “Matter to the Deepest”, Chorzów, Poland, September 1--6, 2019.}%
}
\author{Ievgen Dubovyk$^{a}$, Janusz Gluza$^{a,b}$, Tord Riemann$^{a}$} 

\date{%
{\normalsize    
    $^{a}$Institute of Physics, University of Silesia, Katowice, Poland\\%
    $^{b}$Faculty of Science, University of Hradec Kr\'alov\'e, Czech Republic%
}}

\maketitle

\begin{abstract}
The status of numerical evaluations of Mellin-Barnes integrals is discussed, in particular the application of the quasi-Monte Carlo integration package \texttt{QMC} to the efficient calculation of multi-dimensional integrals. 
\end{abstract}
%\PACS{02.70.Wz,12.15.Lk,12.38.Bx}
 
\section{Introduction}
Recently the Mellin-Barnes (MB) method \cite{Smirnov:2006rya,Czakon:2005rk,Gluza:2007rt,Smirnov:2009up,Gluza:2010rn,Dubovyk:2016ocz,ambrewww,Usovitsch:2018qdk,Dubovyk19} has been applied, together with the sector decomposition method  \cite{Hepp:1966eg,Binoth:2000ps,Smirnov:2013eza,Borowka:2017idc}, to the numerical calculation of two-loop Feynman integrals needed in the  determination of electroweak precision observables (EWPOs, for definitions and physics aspects see e.g. \cite{Blondel:2018madAPP})
in the Z-boson decay \cite{Dubovyk:2016aqv,Dubovyk:2018rlg,Dubovyk:2019szj}.
The $Z$ resonance is formed by electron-positron collisions at center-of-mass energy around $91$ GeV. 
Up to $5 \times 10^{12}$ $Z$-boson decays are planned to be 
observed at projected future $e^{+} e^{-}$ machines (ILC, CEPC, FCC-ee), when  
running at the $Z$-boson resonance \cite{Baer:2013cma,Gomez-Ceballos:2013zzn,%
cepcAPP,%
%CEPC-SPPCStudyGroup:2015csa,%
Fujii:2019zll,Abada:2019zxq,Blondel:2019qlh}.
These statistics are several orders of magnitude larger than that at LEP and would lead to very accurate experimental measurements of EWPOs -- if the systematic experimental errors can be kept appropriately small too. This, in turn, means, that theoretical predictions must be also very precise, of the order of 3- to 4-loop EW and QCD effects \cite{Blondel:2018madAPP}.

\section{Numerical integration of Mellin-Barnes integrals: transition to the Minkowskian region}
Omitting details of the construction of Mellin-Barnes representations, the final 
form of MB integrals suited for numerical integrations can be represented as follows:
\begin{equation}
I = \frac{1}{(2\pi i)^r} \int\limits_{-i \infty + z_{10}}^{+i \infty + z_{10}} \dots \int\limits_{-i \infty + z_{r0}}^{+i \infty + z_{r0}} 
 \underset{i}{\overset{r}{\Pi}} \; dz_i \;
 f_S(Z) \;  
 \frac{\prod_j \; \Gamma(\varLambda_j)}{\prod_k \; \Gamma(\varLambda_k)} \; f_{\psi}(Z).
 \label{MBIntgenForm}
\end{equation} 
In this expression, the integration goes along paths parallel to imaginary axes 
and the positions of contours are fixed by $z_{ i0}$. The Gamma functions depend on linear
combinations of integration variables and some integer numbers. The function $f_{S}(Z)$
depends on ratios of kinematic parameters and internal masses, raised to some powers which are also
linear combinations of integration variables. The part $f_{\psi}(Z)$ may depend on
polygamma functions and constants like the Euler–Mascheroni constant $\gamma_E$; it is equal to one if the
corresponding Feynman integral has no $1/\epsilon^i$ poles. We call the ratio of the gamma-type
functions, $f_{S}(Z)$ and $f_{\psi}(Z)$ the core, head, and tail of the MB integral, 
respectively.

In order to understand the problems which appear due the transition to Min\-kowskian kinematics, 
one has to study the asymptotic behavior of integrands.

The core of MB integrals in case of integration contours parallel to the imaginary axes, 
namely $z_i = z_{i0} + i t_i$, is a smooth function.
Its asymptotic behavior in generalized spherical coordinates can be written as
\begin{equation}
  \frac{\prod_j \; \Gamma(\varLambda_j)}{\prod_k \; \Gamma(\varLambda_k)}
  \xlongrightarrow[|z_i| \rightarrow \infty]{r \rightarrow \infty}
  \frac{e^{- \beta r}}{r^{\alpha}}, \,\,\, \beta = \beta(\theta) \geq \pi,
  \,\,\, \alpha = \alpha(z_{i0}).
\end{equation}

The asymptotic behavior of the tail $f_{\psi}(Z)$ can be omitted.

The head $f_{S}(Z)$ of the MB integral defines the most important asymptotic properties. Let us consider a typical $f_{S}(Z)$ which appears for example in MB integrals for 2-loop form-factors with one or more equal internal masses, 
\begin{equation}
 \left( \frac{m^2}{-s} \right)^z = e^{z \ln (- \frac{m^2}{s} + i \, \delta )}  \longrightarrow e^{i \, t \ln \frac{m^s}{s}} e^{- \pi t}, \,\,\, s>0.
\end{equation}
The infinitesimal parameter $\delta$ in $s \rightarrow s + i \delta \, (s > 0)$ for the Minkowskian case comes from the causality principle and defines the correct sheet of the Riemann surface for the logarithm and the corresponding sign of the imaginary part of the integral.

As one can see, an oscillating behavior of the integrand is a natural feature of MB integrals. The main difficulties in Minkowskian kinematics come with the factor $e^{-\pi t}$. For certain classes of integrals this factor cancels the $e^{- \beta r}$ part of the core along some direction or in some sector of the integration space, and the integrand tends to 0 only as fast as $1 / r^{\alpha}$. In general, such a behavior cannot easily be stabilized such that sufficiently accurate results are in reach.
{One should stress that the overall exponential damping factor in some cases can be restored by deforming the path of integration \cite{Freitas:2010nx}.
Alternatively, here we focus on a direct integration as a new, more general approach.}

In practice, for numerical integrations some external library like \texttt{CUBA} \cite{Hahn:2004fe} is used. Usually, the integration over infinite intervals requires their transformation into finite ones; for example in \texttt{CUBA} it is the interval $[0,1]$.
In the package \texttt{MB.m} \cite{Czakon:2005rk} such transformation is done in the following way:
\begin{equation}
 t_i \rightarrow \ln \left( \frac{x_i}{1 - x_i} \right), \;\;\;
       dt_i \rightarrow \frac{d x_i}{x_i(1 - x_i)}.
\label{lntransform}       
\end{equation} 
This type of transformation leads to an integrable endpoint singularity and makes accurate integrations quite difficult.
As an alternative one can transform the integration interval $(- \infty, \infty)$ into $[0,1]$ in a different way:
\begin{equation}
t_i \rightarrow \tan \left( \pi(x_i - \frac12) \right), \;\;\; 
d t_i \rightarrow \frac{\pi d x_i}{\cos^2\left(\pi(x_i - \frac12)\right)},
\label{tantransform}
\end{equation}
without the appearance of endpoint singularities. For more technical details see \cite{Dubovyk:2016ocz, Usovitsch:2018qdk, Dubovyk19}.

\begin{figure}[!t]
\begin{center}
\includegraphics[scale = 0.75]{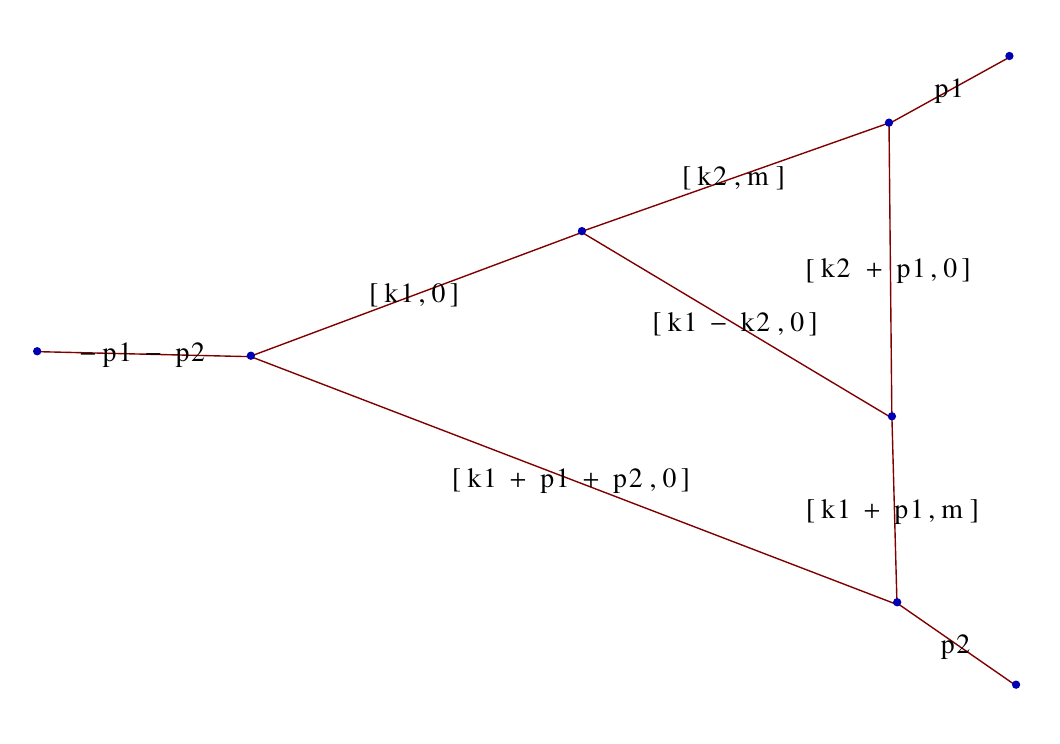}
\end{center}
\caption{An example of a 2-loop vertex diagram with  $(p_1+p_2)^2 = s$ and $p_1^2=p_2^2=0$.
The numerical precision obtained with the MB method is discussed in the text. The diagram is drawn by the \texttt{PlanarityTest.m} package \cite{Bielas:2013v12}.}
\label{VertexDiag}
\end{figure}

As an example of practical calculations, we present here results obtained for the 2-loop vertex diagram shown in Fig.~\ref{VertexDiag}. 
The MB representation for this diagram is three-dimensional:
\begin{align}
I = \frac{1}{(2 \pi i)^3} \frac{1}{s^2}  
 \int\limits_{-i \infty - \frac{47}{37}}^{i \infty - \frac{47}{37}} dz_1 
 \int\limits_{-i \infty - \frac{44}{211}}^{i \infty - \frac{44}{211}} dz_2
 \int\limits_{-i \infty - \frac{176}{235}}^{i \infty - \frac{176}{235}} dz_3
 \left(\frac{m^2}{-s} \right)^{z_1} \nonumber \\
  \Gamma(-1 - z_1) \Gamma(2 + z_1)  
 \Gamma(-1 - z_{12}) \Gamma(-z_2) \Gamma^2(1 + z_{12} - z_3) \Gamma(1 + z_3) \nonumber \\ 
  \Gamma(-z_3)  \Gamma^2(-z_1 + z_3) \Gamma(-z_{12} + z_3) 
   / \Gamma(-z_1) \Gamma(1 - z_2) \Gamma(1 - z_1 + z_3)
   .
\label{3dimEx11} 
\end{align}
The diagram has also an analytical solution \cite{Aglietti:2004tq} which makes it ideal for a non-trivial testing and comparison of different numerical techniques.

\begin{table}[!b]
\caption{\label{tab:NumTabI2}
Numerical results for the integral Eq.~(\ref{3dimEx11}) for $s = m^2 = 1$. \texttt{AB} - analytical solution \cite{Aglietti:2004tq}. \texttt{MB}$1$ to \texttt{MB}$8$ -- numerical integration of the MB integrals with different integration routines and transformations of the infinite integration region as described in the text.} 
\centering
\begin{tabular}{|l|ll|l|} \hline
\rule{0pt}{2.3ex}\texttt{AB}      & $-1.{\bf199526183135}$ & $+5.{\bf567365907880} i$ &   \\ \hline 
\rule{0pt}{2.3ex}\texttt{MB}$1$   & $-1.{\bf19952}5259137$ & $+5.{\bf56736}7419371 i$ & Cuhre,  $10^7$, $10^{-8}$  \\ \hline
\rule{0pt}{2.3ex}\texttt{MB}$2$   & $-1.{\bf19952}4318757$ & $+5.{\bf567365}298565 i$ & Cuhre,  $10^7$, $10^{-8}$  \\ \hline
\rule{0pt}{2.3ex}\texttt{MB}$3$   & $-1.{\bf199526}239547$ & $+5.{\bf567365}843910 i$ & Cuhre,  $10^7$, $10^{-8}$  \\ \hline
\rule{0pt}{2.3ex}\texttt{MB}$4$   & $-1.{\bf1995261831}68$ & $+5.{\bf567365907}904 i$ & Cuhre,  $10^7$, $10^{-8}$  \\ \hline
\rule{0pt}{2.3ex}\texttt{MB}$5$   & \texttt{NaN}           &                          & Cuhre,  $10^7$             \\ \hline 
\rule{0pt}{2.3ex}\texttt{MB}$6$   & $-1.{\bf20}4597845834$ & $+5.{\bf567}518701898 i$ & Vegas,  $10^7$, $10^{-3}$  \\ \hline
\multicolumn{4}{|l|}{} \\ \hline
\rule{0pt}{2.3ex}\texttt{MB}$7$   & $-1.{\bf1995}16455248$ & $+5.{\bf5673}76681167 i$ & QMC, $10^6$, $10^{-5}$     \\ \hline
\rule{0pt}{2.3ex}\texttt{MB}$8$   & $-1.{\bf19952}7580305$ & $+5.{\bf56736}7345229 i$ & QMC, $10^7$, $10^{-6}$     \\ \hline
\end{tabular}
\end{table} 

Eq.~(\ref{3dimEx11}) features a cancellation
of the overall damping factor along the line $t_1=-t_2=t, \, t_3=0$.
After linear transforming  $z_2 \rightarrow z_2 - z_1$, the  cancellation can be isolated along the $t_1$-axis ($t_1=t, \, t_2=t_3=0$). Numerical results for both integral versions obtained with different combinations of transformations (\ref{lntransform}) and (\ref{tantransform}) are compared with an analytical solution in Tab.~\ref{tab:NumTabI2}. 
In the table, the label
\texttt{MB}$1$ corresponds to the numerical integration of Eq.~(\ref{3dimEx11}), where the mapping into
the integration interval $[0,1]$ is done by the $\tan$-type of transformation (\ref{tantransform}) for all variables. 
\texttt{MB}$2$ - integration of Eq.~(\ref{3dimEx11}),  $\tan$-mapping for $t_1$ and $t_2$,
$\ln$-mapping (\ref{lntransform}) for $t_3$. 
\texttt{MB}$3$ - Eq.~(\ref{3dimEx11}) after the transformation $z_2 \rightarrow z_2 - z_1$ and with $\tan$-mapping for all variables. 
\texttt{MB}$4$ - Eq.~(\ref{3dimEx11}) after the transformation, $\tan$-mapping for $t_1$ and $\ln$-mapping for the remaining variables.
\texttt{MB}$5$ - Eq.~(\ref{3dimEx11}), $\ln$-mapping for all variables. 
All integrations are done by the \texttt{CUHRE} routine 
of the \texttt{CUBA} library. The maximum number of integrand evaluations allowed was set to $10^7$. 
The absolute error reported by the routine is at the level of $10^{-8}$.
\texttt{MB}$6$ -  the same as \texttt{MB}$5$, but the integration is done by the \texttt{VEGAS} routine \cite{Lepage:1977sw,Lepage:1980dq} with an error estimation of $\sim 10^{-3}$. 
The last two rows \texttt{MB}$7$ and \texttt{MB}$8$ show results for the numerical integration of Eq.~(\ref{3dimEx11}) and $\tan$-mapping for all variables with the newly presented quasi-Monte Carlo library \texttt{QMC} 
\cite{Borowka:2018goh}. 
Numbers in the last column give the maximum number of integrand evaluations and the absolute error.

The instances 
\texttt{MB}$4$ and
\texttt{MB}$5$
%MB4 and MB5 
in Tab.~1 correspond to the integration with \texttt{MB.m}. They have endpoint singularities due to the $\ln$-type of mapping for all variables. The Monte Carlo algorithm implemented in \texttt{VEGAS} can treat such singularities, but with very low accuracy, which in principle correlates with the maximal number of integrand evaluations. The deterministic \texttt{CUHRE} algorithm is less prepared for such singular behavior and falls to the \texttt{NaN} result after some number of integrand evaluations. Cases 
\texttt{MB}$1$ and
\texttt{MB}$4$
%MB1 to MB4 
are non-singular already and reflect different levels of optimization of the asymptotic behavior. The most accurate result was obtained in the 
\texttt{MB}$4$ case. This case requires exact identification of the direction where the cancellation of the overall damping factor takes place and a rotation of integration variables such that this direction is parallel to one of the axes. In practice, that can be a quite nontrivial task, especially for more-dimensional integrals or for more scales. In the case \texttt{MB}$1$, the direction of the cancellation is not identified. The \texttt{MB}$1$ and the $\tan$-type of mapping only fixes the endpoint singularity. One should stress that in all cases \texttt{MB}$1$ to
\texttt{MB}$4$
the error estimation is at the level of $10^{-8}$, but the true number of correct digits is different all the time and doesn't correspond to the error (under)estimation probability returned by the program\footnote{{%
The CUBA library, together with result and absolute error, also returns a probability that the error estimate is not reliable. According to this probability, the error of $10^{-8}$ in the \texttt{MB}$1$ case is reliable and one would expects seven correct digits -- but the number of trusted digits is only five. In the \texttt{MB}$4$ case the error estimation is not trustable and one would expect less than seven correct digits. 
In practice, this is the most accurate result.}}.
That makes \texttt{CUHRE} not truly reliable for such types of integrals, and it was the main motivation to develop the \texttt{MBnumerics} package \cite{Dubovyk:2016ocz, Usovitsch:2018qdk} which is not sensitive to this kind of problem of \texttt{CUHRE}.
In contrast to \texttt{CUHRE}, the \texttt{QMC} library gives a stable error estimation and the requested accuracy can be obtained just by increasing the number of integration points, without any other efforts {such as seeking transformation coefficients to improve the asymptotic behavior of the integrand. The obtained error is bigger than with \texttt{CUHRE} for the same number of integrand evaluations.  This is typical for quasi-MC or Monte Carlo methods and will surpass \texttt{CUHRE} for more dimensional integrals.} 

\section{Conclusions}
Currently, the \texttt{QMC} library is one of the most suitable tools for the numerical integration of MB integrals in the Minkowskian region.
The library shows a linear dependence between the number of integration points and the number of correct digits in the result. {This property makes it more convenient for high-dimensional integrals in contrast to deterministic or pure Monte Carlo algorithms.}
A combination of the appropriate transformation of the infinite integration region into a finite one  with the \texttt{QMC} integrator allows the calculation of a wide class of MB integrals with an acceptable accuracy. All results shown here were calculated in single-thread mode on an Intel i5 3310M mobile CPU within few minutes per case. This fact gives extra room for applications to more complicated problems and for accuracy improvements on more powerful computers.
The integration of \texttt{QMC} into the \texttt{MBnumerics} package in order to get optimal accuracy and speed certainly needs further studies.

\section*{Acknowlegements}
The work was supported partly by the Polish National Science Centre (NCN) under the Grant Agreement 2017/25/B/ST2/01987,  the international mobilities for research activities of the University
of Hradec Kr\'alov\'e,
CZ.02.2.69/0.0 /0.0/16027/0008487, and the COST (European Cooperation in Science and Technology) Action CA16201 PARTICLEFACE.
The participation of T.R. at MTTD2019 was kindly supported by DESY. 

%\bibliographystyle{elsarticle-num}
%\bibliography{yr_refs_update,2loopsreport,phd_dubovyk,MBmethods,2loops,%2loops_ambre2015jg}

\end{document}